\documentclass[aps,prl,twocolumn,superscriptaddress,showpacs]{revtex4}
\usepackage{dcolumn}
\usepackage{bm}
\usepackage{epsfig}

\newcommand{\etal}{{\it et al.}}

\begin{document}

\begin{minipage}{3.4cm}

\hspace*{5.7 in}CLNS~05/1930\\
\hspace*{5.7 in}CLEO~\mbox{05-18}\\
\end{minipage}


\title{\LARGE First Evidence and Measurement \\
of $B_s^{(*)}\overline{B_s}^{(*)}$
Production at the $\Upsilon$(5S)}

\author{M.~Artuso}
\author{C.~Boulahouache}
\author{S.~Blusk}
\author{J.~Butt}
\author{O.~Dorjkhaidav}
\author{J.~Li}
\author{N.~Menaa}
\author{R.~Mountain}
\author{R.~Nandakumar}
\author{K.~Randrianarivony}
\author{R.~Redjimi}
\author{R.~Sia}
\author{T.~Skwarnicki}
\author{S.~Stone}
\author{J.~C.~Wang}
\author{K.~Zhang}
\affiliation{Syracuse University, Syracuse, New York 13244}
% ...
\collaboration{CLEO Collaboration} 
\noaffiliation

\date{August 22, 2005}

\begin{abstract}
We use data collected by the CLEO III detector at CESR to measure
the inclusive yields of $D_s$ mesons as ${\cal{B}}(\Upsilon(5S)\to
D_sX)=(44.7\pm 4.2 \pm 9.9)\%$ and ${\cal{B}}(\Upsilon(4S)\to
D_sX)=(18.1\pm0.5\pm2.8)\%$. From these measurements we make a
model dependent estimate of the ratio of
$B_s^{(*)}\overline{B_s}^{(*)}$ to the total $b\overline{b}$ quark
pair production of $(16.0\pm 2.6 \pm 5.8)$\% at the $\Upsilon$(5S)
energy.
\end{abstract}
\pacs{13.25.Hw, 13.66.Bc} \maketitle

An enhancement in the total $e^+e^-$ annihilation cross-section
into hadrons was discovered at CESR long ago
\cite{CLEOIImeasurement,CUSBmasses} and its mass measured as
10.865$\pm$0.008 GeV. This effect was named the $\Upsilon$(5S)
resonance. Theoretical models \cite{UQM,NQM,CC} predict the
different relative decay rates of the $\Upsilon$(5S) into
combinations of $B^{(*)}\overline{B}^{(*)}$ and $B_s^{(*)}
\overline{B}_s^{(*)}$ where $(*)$ indicates the possible presence
of a $B^*$ meson. This original $\sim$116 pb$^{-1}$ of data failed
to reveal if $B_s$ mesons were produced. It is important to check
the predictions of these and other models; furthermore, $e^+e^-$
``B factories" could exploit a possible $B_s$ yield here as they
have done for $B$ mesons on the $\Upsilon$(4S).

In this Letter we examine $D_s$ yields because in a simple spectator
model the $B_s$ decays into the $D_s$ nearly all the time. Since the
$B\to D_s X$ branching ratio has already been measured to be
$(10.5\pm 2.6\pm 2.5)\%$ \cite{PDG}, we expect a large difference
between the $D_s$ yields at the $\Upsilon$(5S) and the
$\Upsilon$(4S) that can lead to an estimate of the size of the
$B_s^{(*)}\overline{B_s}^{(*)}$ component at the $\Upsilon$(5S).
When we discuss the $\Upsilon$(5S) here, we mean any production
above what is expected from continuum production of quarks lighter
than the $b$ at an $e^+e^-$ center-of-mass energy of 10.865 GeV. The
CLEO III detector is equipped to measure the momenta and directions
of charged particles, identify charged hadrons, detect photons, and
determine with good precision their directions and energies. It has
been described in detail previously in references \cite{CLEODR} and
\cite{RICH}.


In this analysis we use 0.42 fb$^{-1}$ of integrated luminosity
taken at the $\Upsilon$(5S) peak in Feb. 2003. We also use 6.34
fb$^{-1}$ of integrated luminosity collected on the $\Upsilon$(4S)
and 2.32 fb$^{-1}$ of data taken in the continuum 40 MeV in
center-of-mass energy below the $\Upsilon$(4S). These data were
accumulated between Aug. 2000 and June 2001. The detector hardware
wasn't changed over the entire time period. Efficiencies are
carefully monitored and did not change measurably between data sets.

We look for $D_s$ candidates through the reconstruction of three
charged tracks in hadronic events via the
$D_s^+\rightarrow\phi\pi^+$ decay mode. Here and elsewhere in this
paper mention of one charge implies the same consideration for the
charge-conjugate mode. Requiring the Fox-Wolfram shape parameter
$R_2$ \cite{Fox} to be less than $0.25$ suppresses continuum
background events which are less isotropic than $b$-quark events.

Pairs of oppositely charged tracks were considered candidate decay
products of a {$\phi$} if at least one of the tracks is identified
as a kaon, and if the invariant mass of the $K^+K^-$ system is
within $\pm 10$ $\rm MeV/c^2$ of the nominal {$\phi$} mass. A third
track was combined with the $K^+K^-$ system to form a $D_s$
candidate without using particle identification.

The Ring Imaging Cherenkov (RICH) of the CLEO III detector is used
for track momenta larger than 0.62 GeV/c. Information on the angle
of the detected Cherenkov photons is translated into a Likelihood of
a given photon being due to a particular particle analyzed with a
specific mass hypothesis. Contributions from all photons associated
with a particular track with one mass hypothesis are then summed to
form an overall Likelihood denoted as ${\cal L}_i$ for each ``$i$"
particle hypothesis.

To utilize the information on the ionization loss in the drift
chamber of the CLEO III detector, dE/dx, we calculate the
differences between the expected and the observed ionization losses
divided by the error for the pion and kaon hypotheses, called
$\sigma_{\pi}$ and $\sigma_{K}$.

We use both RICH and dE/dx information in the following manner: (a)
If neither RICH nor dE/dx information is available, then the track
is accepted. (b) If dE/dx is available and RICH is not, then we
insist that kaon candidates have
$PID_{dE}=\sigma_{\pi}^2-\sigma_{K}^2> 0.$ (c) If RICH information
is available and dE/dx is not available, then we require that
$PID_{RICH}=-2\log({\cal L}_{\pi})+2\log({\cal L}_K)>0$ for kaons.
(d) If both dE/dx and RICH information are available, we require
that $(PID_{dE}+PID_{RICH})>0$ for kaons.

To suppress combinatoric backgrounds, we take advantage of the
polarization of the $\phi$ as it is a vector particle while the
other particles in this decay are spinless. The expected
distribution from real $\phi$ decays varies as $\cos^2\theta_h$,
where $\theta_h$ is the angle between the $D_s$ and the $K^+$
momenta measured in the $\phi$ rest frame while combinatoric
backgrounds tend to be flat. Thus, we require $|\cos\theta_h|$ to be
larger than $0.3$.



\begin{figure}[htb]
\centerline{\epsfig{figure=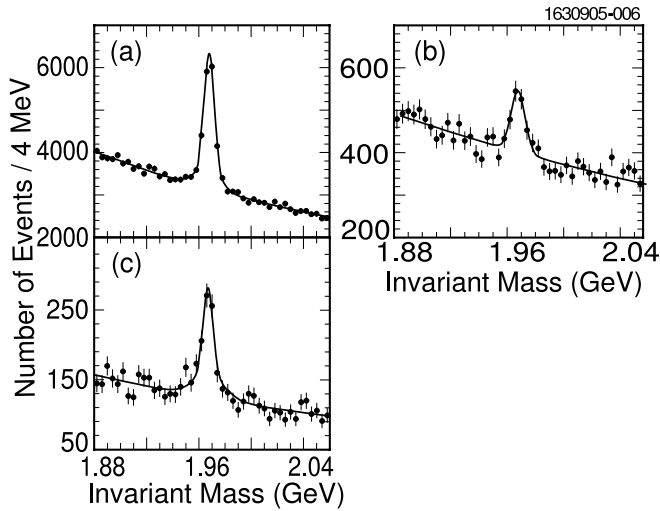,width=3.4in}}
  \caption{\label{4s-5s-ds-inv-mass} The invariant mass distributions of the $D_s$ candidates
  with $x<$ 0.5 from: (a) the $\Upsilon$(4S) on-resonance
  data (b) the continuum below the $\Upsilon$(4S)
resonance data (c) the $\Upsilon$(5S) on-resonance data.}
\end{figure}

For $\phi\pi^+$ combinations satisfying the previous requirements,
we look for $D_s$ candidates having a momentum less than half of the
beam energy. Instead of momentum we choose to work with the variable
$x$ which is the $D_s$ momentum divided by the beam energy, to
remove differences caused by the change of the beam energies between
continuum data taken just below the $\Upsilon$(4S), at the
$\Upsilon$(4S) and at the $\Upsilon$(5S). The $\phi\pi$ invariant
mass distributions for $x < 0.5$ are shown in
Fig.~\ref{4s-5s-ds-inv-mass}.

We fit the invariant mass of the $\phi\pi^{\pm}$ candidates in 10
different $x$ intervals (from 0 to 0.5) for data taken at the
$\Upsilon$(4S) peak, at the continuum below the $\Upsilon$(4S) and
at the $\Upsilon$(5S) peak.

The invariant mass distribution in each $x$ interval of the
$\Upsilon$(4S) data set was fit to a Gaussian signal shape and a
linear background. The width of each Gaussian was allowed to float.
The corresponding distributions at the other energies were similarly
fit, but with the corresponding Gaussian widths fixed to those
determined at the $\Upsilon$(4S). The raw $D_s$ yields are
listed in the second, third and fourth columns of
Table~\ref{tab:Ds4S}.

The number of $D_s$ candidates is determined by subtracting the
scaled four-flavor ($u$, $d$, $s$ and $c$ quark) continuum data
below the $\Upsilon$(4S) from the $\Upsilon$(4S) and from the
$\Upsilon$(5S) data. To determine the scale factors, $S_{nS}$, we
account for both the ratio of luminosities and the $s$ dependence of
the continuum cross section using
\begin{equation}
S_{nS}={\frac{L_{nS}}{L_{cont}}}\cdot{
\left({\frac{E_{cont}}{E_{nS}}}\right)^2}
\end{equation}
where $L_{nS}$, $L_{cont}$, $E_{nS}$ and $E_{cont}$ are the
collected luminosities and the center-of-mass energies at the
$\Upsilon$(nS) and at the continuum below the $\Upsilon$(4S). We
find: $S_{4S}={2.712 \pm 0.001 \pm 0.043}$ and $S_{5S}={(17.15 \pm
0.01 \pm 0.24)}\cdot{10^{-2}}$.

The second (systematic) error on these scale factors is determined
by using the number of charged tracks in the $0.6<x<0.8$ interval.
The lower limit is determined by the maximum value tracks from
$B\overline{B}$ events can have, including smearing due to the
measuring resolution, and the upper limit is chosen to eliminate
radiative electromagnetic processes. Since the tracks should be
produced only from continuum events, we suppress beam-gas and
beam-wall interactions, photon pair and $\tau$ pair events using
strict cuts on track multiplicities, event energies and event
shapes. (Since particle production may be larger at the higher
$\Upsilon (5S)$ energy than the continuum below the $\Upsilon (4S)$,
we apply a small multiplicative correction of (0.6$\pm$1.1)\%, as
determined by Monte Carlo simulation to the relative track yields.)
We find that the scale factors using this track counting method are
2.668$\pm$0.007, and $(17.085\pm 0.207)\cdot 10^{2}$, for $S_{4S}$
and $S_{5S}$, respectively, and use the difference as the systematic
error.

The total number of hadronic events above four-flavor continuum are
$N^{Res}_{\Upsilon(4S)}$ equals $(6.43\pm 0.01\pm
0.41)\cdot{10^{6}}$ and $N^{Res}_{\Upsilon(5S)}$ equals $(0.130\pm
0.001 \pm 0.022)\cdot{10^{6}}$. The $6.4\%$ and $17.5\%$ systematic
errors here are due to the $1.6\%$ and $1.4\%$ systematic errors on
$S_{4S}$ and $S_{5S}$ scale factors respectively.

The branching ratio of $\Upsilon(nS)\to D_sX$ in each $i$-th $x$
interval is given by
\begin{eqnarray}
\lefteqn{{\cal{B}}^{i}(\Upsilon(nS)\to D_sX)=}\nonumber\\
&&\!\!\!\!\!\!{\frac{1}{N^{Res}_{\Upsilon(nS)}\cdot{{\cal{B}}(D_{s}\to
\phi \pi)}\cdot{{\cal{B}}(\phi\to
K^+K^-)}}}~\left({\frac{N^{i}_{\Upsilon(nS)}}
{\epsilon^{i}}}\right),~\label{eq:RSol4}\nonumber\\
\end{eqnarray}
where $N^{i}_{\Upsilon(nS)}$ are the continuum subtracted on
resonance $D_s$ yields. ${\cal{B}}(\phi\to K^+K^-)$ is taken as
49.1\% \cite{PDG}. The reconstruction efficiency $\epsilon^{i}$ is
taken to be the same at both resonances. This is reasonable because
our tracking and particle identification efficiencies are carefully
monitored and did not change significantly between data sets.
Specifically, our Monte Carlo simulations of the $D_s$
reconstruction efficiencies includes time dependent effects of dead
channels and individual hit efficiencies in both the tracking and
RICH systems. A comparison of the simulations at both energies shows
changes in the reconstruction efficiency between the two energies of
$<$2\%.

\begin{figure}[htbp]
 \centerline{\epsfig{figure=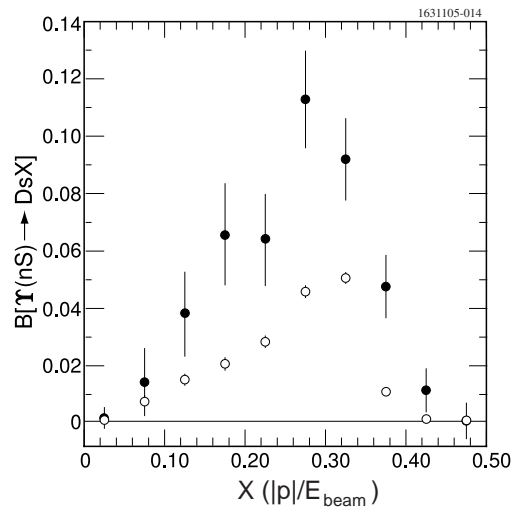, width=2.6in}}
  \caption{\label{Dsyield} Branching rate as a function of $x$ from
  $\Upsilon$(5S) decays (filled circles) and from $\Upsilon$(4S) decays
  (open circles).}
\end{figure}

 The
results are listed in Table~\ref{tab:Ds4S}. We show in
Fig.~\ref{Dsyield} the $x$ distribution of the inclusive $D_s$
yields from $\Upsilon$(4S) and $\Upsilon$(5S) decays, continuum
subtracted and efficiency corrected.

\begin{table*}[htb]
\begin{center}
\begin{tabular}{ccccccccc}
\hline\hline $x^{i}$($\frac{|p|}{Ebeam}$)~ &~ ON $\Upsilon$(4S) & ON
$\Upsilon$(5S) & Continuum &
$N^{i}_{\Upsilon(4S)}$ & $N^{i}_{\Upsilon(5S)}$ & $\epsilon^{i}(\%)$ &~ $B^{i}_{4S}$(\%) &~  $B^{i}_{5S}$(\%) \\
 \hline
0.00-0.05 & $44.4\pm15.7$   & $1.0\pm3.2$    &~ $0.0\pm0.0$   &~  $44.4\pm15.7 $   &~ $1.0\pm3.1$       &~  $28.9 $  &~   $0.11\pm0.04 $ &~  $0.1\pm0.4 $\\
0.05-0.10 & $317.6\pm39.6$  & $13.3\pm8.1$   &~ $20.7\pm12.0$ &~  $261.4\pm51.2 $  &~ $9.7\pm8.3$   &~  $23.9 $  &~   $0.8\pm0.2 $ &~ $1.4\pm1.2 $\\
0.10-0.15 & $583.6\pm53.9$  & $30.4\pm10.4$  &~ $21.6\pm15.3$ &~  $524.9\pm68.1 $  &~ $26.7\pm10.7$ &~  $24.7 $  &~   $1.5\pm0.2 $ &~ $3.8\pm1.5 $ \\
0.15-0.20 & $845.5\pm59.0$  & $54.4\pm13.0$  &~ $41.7\pm18.5$ &~  $732.3\pm77.5 $  &~ $47.2\pm13.3$ &~  $25.4 $  &~   $2.1\pm0.2 $ &~ $6.5\pm1.8 $ \\
0.20-0.25 & $1206.4\pm60.6$ & $57.6\pm12.7$  &~ $40.2\pm18.3$ &~  $1097.4\pm78.2 $ &~ $50.7\pm13.0$ &~  $27.7 $  &~   $2.8\pm0.2 $ &~ $6.4\pm1.7 $ \\
0.25-0.30 & $2028.6\pm63.8$ & $104.1\pm14.0$ &~ $70.3\pm18.0$ &~  $1838.0\pm80.3 $ &~ $92.0\pm14.3$ &~  $28.6 $  &~   $4.6\pm0.2 $ &~ $11.3\pm1.8 $ \\
0.30-0.35 & $2233.7\pm60.7$ & $86.7\pm12.1$  &~ $57.0\pm16.2$ &~  $2079.2\pm74.9 $ &~ $76.9\pm12.4$ &~  $29.4 $  &~   $5.0\pm0.2 $ &~ $9.2\pm1.5 $ \\
0.35-0.40 & $660.8\pm37.9$  & $53.8\pm9.4$   &~ $75.0\pm14.5$ &~  $457.4\pm54.6 $  &~ $41.0\pm9.7$  &~  $30.4 $  &~   $1.1\pm0.1 $ &~ $4.7\pm1.1 $ \\
0.40-0.45 & $233.5\pm25.9$  & $22.6\pm6.7$   &~ $73.4\pm12.9$ &~  $34.3\pm43.3 $   &~ $10.1\pm7.0$  &~  $31.4 $  &~   $0.1\pm0.1 $ &~ $1.1\pm0.8 $ \\
0.45-0.50 & $245.8\pm22.2$  & $14.8\pm5.6$   &~ $86.0\pm12.1$ &~  $12.6\pm39.5 $   &~ $0.1\pm6.0$   &~  $32.4 $  &~   $0.03\pm0.09 $ &~ $0.0\pm0.6 $ \\
\hline\hline
\end{tabular}
\end{center}
\caption{The $x$ dependent $D_s$ yields from the $\Upsilon$(nS)
data, the continuum below the $\Upsilon$(4S), the $\Upsilon$(nS)
continuum subtracted data, $N^{i}_{\Upsilon(nS)}$, the efficiency
$\epsilon^i$, and the partial branching ratios
$B^i_{nS}$=$\Upsilon(nS)\to D_s X$, for $ns$ equal to 4S and 5S. The
errors are statistical only.} \label{tab:Ds4S}
\end{table*}

The total production rate is found by summing the partial production
rates. The product of the $D_s$ production rate at the
$\Upsilon$(4S) and the ${\cal{B}}(D_s\to\phi\pi)$ is
\begin{equation}
{\cal{B}}(\Upsilon(4S)\to D_sX)\cdot{\cal{B}}(D_s\to\phi\pi)=
(8.0\pm 0.2 \pm 0.9)\cdot{10^{-3}}~\label{eq:rate4}
\end{equation}
which is in a good agreement with previous measurements
\cite{PDG}, while at the $\Upsilon$(5S)
\begin{equation}
{\cal{B}}(\Upsilon(5S)\to D_sX)\cdot{\cal{B}}(D_s\to\phi\pi)=
(19.8\pm 1.9 \pm 3.8)\cdot{10^{-3}}~~.\label{eq:rate5}
\end{equation}
Many systematic errors cancel in the ratio of decay rates. Thus
\begin{equation}
{{{\cal{B}}(\Upsilon({\rm 5S})\to D_s X)} \over
{{\cal{B}}(\Upsilon({\rm 4S})\to D_s X)}} = 2.4\pm
0.3^{+0.6}_{-0.3}~,
\end{equation}
directly demonstrates, at 5.6 standard deviation significance, a
much larger yield of $D_s$ at the $\Upsilon$(5S) than at the
$\Upsilon$(4S).

We use ${\cal{B}}(D_s\to\phi\pi^+)= (4.4 \pm 0.5)\%$, which is the
weighted average of the $( 3.6 \pm 0.9 )\%$ PDG value \cite{PDG}
and the recent measured value of $( 4.8 \pm 0.6 )\%$
\cite{DsNewMeas}, although the latter value is at the 90\% c.l.
upper limit found previously \cite{Muheim}. We find
\begin{equation}
{\cal{B}}(\Upsilon(4S)\to
D_sX)=(18.1\pm0.5\pm2.8)\%,~~\label{eq:4stoDsMeasured}
\end{equation}
and consequently:
\begin{equation}
{\cal{B}}(B\to
D_sX)=(9.0\pm0.3\pm1.4)\%~.~~\label{eq:BtoDsMeasured}
\end{equation}

In addition, we find
\begin{equation}
{\cal{B}}(\Upsilon(5S)\to D_sX)=(44.7\pm 4.2 \pm
9.9)\%~.~\label{eq:5stoDsMeasured}
\end{equation}


From these results, we estimate the size of
$B_s^{(*)}\overline{B}_s^{(*)}$ component at the $\Upsilon$(5S) in
a model dependent manner.  Here we start with the knowledge that
an equal admixture of $B^o$ and $B^+$ mesons decay into the sum of
$D^o$ and $D^+$ mesons roughly 100\% of the time \cite{PDG}. Thus
we expect $B_s$ mesons to decay into $D_s$ mesons also about 100\%
of the time. In what follows we estimate our own theoretical
corrections to this number.
\begin{figure}[htb]
 \centerline{\epsfig{figure=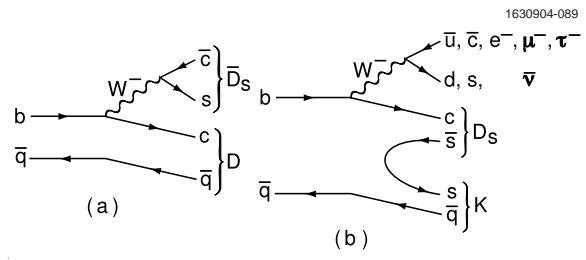,width=3.in}}
  \caption{\label{BtoDs} Dominant decay diagrams for a $B$ meson into $D_s$ mesons
  ($q$ is either a $u$ or $d$ quark).}
\end{figure}

We know that the branching fraction ${\cal{B}}(B\to D_sX)=(9.0\pm
0.3\pm 1.4)\%$ comes either from the $W^-\to \overline {c}s$
process, shown in Fig.~\ref{BtoDs}(a), or from the $b\to c$ piece if
it manages to create an $s\overline{s}$ pair through fragmentation,
see Fig.~\ref{BtoDs}(b).

\begin{figure}[htb]
 \centerline{\epsfig{figure=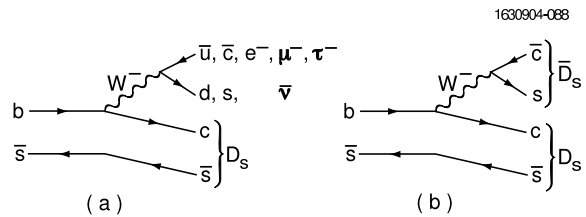,width=3.in}}
  \caption{\label{BstoDs} Dominant decay diagrams for a $B_s$ meson into $D_s$ mesons.}
\end{figure}

Similarly, the production of $D_s$ mesons from $B_s$ decay arises
from two dominant processes. Fig.~\ref{BstoDs}(a) shows the simple
spectator process that is expected to produce $D_s$ mesons nearly
all the time; here the primary $b\to c$ transition has the charm
quark pairing with the spectator anti-strange quark.
Fig.~\ref{BstoDs}(b) shows the subset of process (a) where $W^-\to
\overline {c}s$ and these two quarks form a color singlet pair. The
chances of this occurring should be similar to the chance of getting
an upper-vertex $D_s$ in $B$ decay (Fig.~\ref{BtoDs}(a)), i. e. a
$D_s$ along with a $D$.

We can use data to help estimate the size of these processes. First
let us consider the diagram shown in Fig.~\ref{BstoDs}(a). The
nearly 100\% probability that this process will produce $D_s$ mesons
is reduced if the $c\overline{s}$ pair fragments into a kaon plus a
$D$ instead of a $D_s$ by producing an additional $u\overline{u}$ or
$d\overline{d}$ pair. We don't actually know the size of this
fragmentation, though it's clear that producing a light
quark-antiquark pair ($d\overline{d}$ or $u\overline{u}$) is easier
than $s\overline{s}$. We estimate that the reduction in $D_s$ yield
due to this fragmentation is a $(-15\pm10)\%$ effect. Next we
estimate the size of the process depicted in Fig.~\ref{BstoDs}(b).
The $B\to D D_s$ modes have branching fractions that sum to about
$5\%$. There are some additional decays due to $B\to D^{**} D_s$ and
$B\to D D_{sJ}^{(*)}$ decays, that also contribute $D_s$ mesons. We
add these and estimate an extra ($7\pm3)\%$ of $D_s$ mesons in $B_s$
decays produced by diagram  Fig.~\ref{BstoDs}(b). Taking into
account all these contributions, we derive a model dependent
estimate of $(100+7-15)\%=92\%$. Therefore, we use ${\cal{B}}(B_s\to
D_sX)=(92\pm11)\%$.

We can estimate now the fraction of the $\Upsilon$(5S) that decays
into $B_s^{(*)}\overline{B}_s^{(*)}$, which we denote as $f_s$. The
$D_s$ yields at the $\Upsilon(5S)$ come from two sources, $B$ and
$B_s$ mesons. The equation linking them is

\begin{eqnarray}
\lefteqn{{\cal{B}}(\Upsilon(5S)\to D_s X)
{\cal{B}}(D_s\to\phi\pi^+)/2=}\nonumber\\
&& f_s\cdot {\cal{B}}(B_s\to D_sX){\cal{B}}
(D_s\to\phi\pi^+)\nonumber\\
&&+\frac{(1-f_s)}{2} \cdot{{\cal{B}}(\Upsilon(4S)\to D_s X)}
{\cal{B}}(D_s\to\phi\pi^+)~,\nonumber\\
\end{eqnarray}
where the product branching fractions ${\cal{B}}(\Upsilon(5S)\to D_s
X)\cdot{{\cal{B}}(D_s\to\phi\pi^+)}$ and ${\cal{B}}(\Upsilon(4S)\to
D_s X)\cdot{{\cal{B}}(D_s\to\phi\pi^+)}$ are given by equations
\ref{eq:rate5} and \ref{eq:rate4} respectively. Therefore, at the
$\Upsilon$(5S) energy, we obtain the $B_s^{(*)}\overline{B_s}^{(*)}$
ratio to the total $b\overline{b}$ quark pair production above
four-flavor ($u$, $d$, $s$ and $c$ quarks) continuum of
\begin{equation}
f_s={\cal{B}}(\Upsilon(5S)\to
B_s^{(*)}\overline{B}_s^{(*)})=(16.0\pm 2.6\pm 5.8)
\%~\label{eq:5stoBsEstimated}
\end{equation}


The systematic errors in this analysis are dominated by the $1.6\%$
relative error on $S_{4S}$ and $1.4\%$ on $S_{5S}$ scale factors
which contribute large components ($6.4\%$ and $17.5\%$) to the
error on the number of hadronic events above continuum at the
$\Upsilon$(4S) and $\Upsilon$(5S). There is also a contribution from
the $11.3\%$ error on the $B_s\to D_s X$ branching fraction estimate
and a contribution from the $11\%$ error on the absolute
$D_s\to\phi\pi$ branching fraction. An additional component comes
from a $6.4\%$ error on the $D_s$ detection efficiency, which
includes a $2\%$ error on the tracking efficiency and a $2\%$ error
on the particle identification, both per track. We also have $5\%$
error on the yields due to the fitting method. The total systematic
error is obtained by summing all entries in quadrature.

In conclusion, we have measured the inclusive yields of $D_s$ mesons
as ${\cal{B}}(\Upsilon(5S)\to D_sX)=(44.7\pm 4.2 \pm 9.9)\%$ and
${\cal{B}}(\Upsilon(4S)\to D_sX)=(18.1\pm0.5\pm2.8)\%$. The ratio
\begin{equation}
{{{\cal{B}}(\Upsilon({\rm 5S})\to D_s X)} \over
{{\cal{B}}(\Upsilon({\rm 4S})\to D_s X)}} = 2.4\pm
0.3^{+0.6}_{-0.3}~,
\end{equation}
provides the first statistically significant evidence (5.6$\sigma$)
of substantial production of $B_s$ mesons at the $\Upsilon$(5S)
resonance. Using a model dependent estimate of ${\cal{B}}(B_s\to D_s
X)$, we find that the $B_s^{(*)}\overline{B_s}^{(*)}$ ratio to the
total $b\overline{b}$ quark pair production above the four-flavor
($u$, $d$, $s$ and $c$) continuum at the $\Upsilon$(5S) energy is
\begin{equation}
f_s={\cal{B}}(\Upsilon(5S)\to
B_s^{(*)}\overline{B}_s^{(*)})=(16.0\pm 2.6\pm 5.8)
\%.~\label{eq:5stoBsEstimated2}
\end{equation}

Several phenomenological models predict the decay rates of the
$\Upsilon$(5S) into combinations of $B^{(*)}\overline{B}^{(*)}$ and
$B_s^{(*)} \overline{B}_s^{(*)}$, though here we are only concerned
with the relative $B_s$ fraction $f_s$. The unitarized quark model
estimates \cite{UQM} and the predictions of Martin and Ng \cite{CC}
are about 30\%, both somewhat larger than our measurement. Byers and
Eichten \cite{NQM} present two models both giving $f_s<$ 20\% in
good agreement with our finding.

 We gratefully
acknowledge the effort of the CESR staff in providing us with
excellent luminosity and running conditions. This work was
supported by the National Science Foundation and the U.S.
Department of Energy.

\end{document}